\PassOptionsToPackage{bookmarks=false}{hyperref}
\documentclass[fleqn,11pt]{wlscirep}

\title{Invisible Trojan-horse attack}

\author[1,2,*]{Shihan~Sajeed}
\author[1,3]{Carter~Minshull}
\author[4,$\dag$]{Nitin~Jain}
\author[3,1,2]{Vadim~Makarov}

\affil[1]{Institute for Quantum Computing, University of Waterloo, Waterloo, ON, N2L~3G1 Canada}
\affil[2]{Department of Electrical and Computer Engineering, University of Waterloo, Waterloo, ON, N2L~3G1 Canada}
\affil[3]{Department of Physics and Astronomy, University of Waterloo, Waterloo, ON, N2L~3G1 Canada}
\affil[4]{Department of Physics, Technical University of Denmark, Fysikvej, Kongens Lyngby 2800, Denmark}

\affil[*]{Corresponding author: shihan.sajeed@gmail.com}
\affil[$\dag$]{Corresponding author: nitinj@iitbombay.org}

\usepackage{graphicx}
\usepackage[squaren]{SIunits}
\usepackage{comment}
\usepackage{xcolor}
\usepackage[capitalise]{cleveref}
\crefname{subsection}{subsection}{subsections}
\definecolor{ss_color}{rgb}{0,0,1}

\definecolor{pink}{RGB}{255,0,255}

\newcommand{\ket}[1]{|#1\rangle}

\begin{abstract}
We demonstrate the experimental feasibility of a Trojan-horse attack that remains nearly invisible to the single-photon detectors employed in practical quantum key distribution (QKD) systems, such as Clavis2 from ID~Quantique. We perform a detailed numerical comparison of the attack performance against Scarani-Ac{\' i}n-Ribordy-Gisin (SARG04) QKD protocol at $\boldsymbol{1924~\nano\meter}$ versus that at $\boldsymbol{1536~\nano\meter}$. The attack strategy was proposed earlier but found to be unsuccessful at the latter wavelength, as reported in N.~Jain \textit{et al.,}\ New J.\ Phys.\ 16, 123030 (2014). However at $\boldsymbol{1924~\nano\meter}$, we show experimentally that the noise response of the detectors to bright pulses is greatly reduced, and show by modeling that the same attack will succeed. The invisible nature of the attack poses a threat to the security of practical QKD if proper countermeasures are not adopted.
\end{abstract}

\begin{document}
\flushbottom
\maketitle
\thispagestyle{fancy}

\section*{Executive summary}

\noindent A previous study in 2014 proposed a Trojan-horse attack against Clavis2 receiver (Bob) module; however the attack fell short of the performance level needed to breach the system security -- by a large margin of roughly 100 times. Our present study shows that if an attacker resorts to using a longer wavelength ($> 1900~\nano\meter$) not ordinarily used in telecommunication, the same attack may breach the security. Although a complete eavesdropping apparatus is still quite challenging to build, it might be possible with today's or near-future technology. To prevent this, we have recommended the manufacturer to install a wavelength filter, which is a simple fiber-optic component that can be added just outside the installed system without having to recall it to the factory. For customers using ID~Quantique's QKD products for critical data protection, we recommend that they inquire the manufacturer about this upgrade at the next convenient opportunity, such as a scheduled on-site maintenance. Not every installed system requires this upgrade: some systems are using protocols not vulnerable to this attack, and some may already have the wavelength filter included as part of network configuration. Since QKD cannot be attacked retroactively, security of customers' historical network transmissions is not affected by this study.

\section*{Introduction}\label{sec:intro}

\noindent Quantum cryptography allows two parties, Alice and Bob, to obtain random but correlated sequences of bits by exchanging quantum states~\cite{bennett1984,gisin2002,scarani2009}. The bit sequences can then be classically processed to get shorter but secret keys. The security of the key relies on the fact that an adversary Eve cannot eavesdrop on the exchange without introducing errors noticeable to Alice and Bob. This constitutes a solution to the problem of key distribution in cryptography, and is better known as quantum key distribution (QKD). 

The security of keys distributed over the `quantum channel' connecting Alice and Bob can be validated by a theoretical security proof. If the amount of errors observed by the two parties exceed a certain threshold, they abort the QKD protocol. Conversely, if the incurred quantum bit error rate (QBER) is below the abort threshold $Q_{\rm abort}$, the protocol guarantees that Eve cannot know the secret key, except with a vanishingly small probability~\cite{scarani2009}. 

However, due to discrepancies between theory and practice, the operation of the QKD protocol may be manipulated by Eve in order to gain information about the key without introducing too many errors. Such discrepancies can arise due to imperfections in the physical devices used in the implementation and/or incorrect assumptions in the theoretical security proofs~\cite{scarani2009,makarov2011,Scarani2014}. The field of `quantum hacking' investigates practical QKD implementations to find such theory-practice deviations, demonstrate the resultant vulnerability via proof-of-principle attacks, and propose countermeasures to protect Alice and Bob from Eve. Over the years, many vulnerabilities have been discovered and attacks have been proposed and demonstrated on both commercial and laboratory QKD systems; see refs.~\citen{Lo2014,jain2016,liang2014} for reviews. In most cases, it was shown that under attack conditions, the QBER $Q \leq Q_{\rm abort}$ but Eve's knowledge of the secret key was substantially larger than the predictions of the security proof.

In the so-called Trojan-horse attack~\cite{gisin2006} (introduced as a `large pulse attack' a few years before~\cite{vakhitov2001}), Eve probes the properties of a component inside Alice or Bob by sending in a bright pulse and analyzing a suitable back-reflected pulse. This attack was recently demonstrated~\cite{jain2014} with the intention to breach the security of the Scarani-Ac\'in-Ribordy-Gisin QKD protocol (SARG04)~\cite{scarani2004} running on the commercial QKD system Clavis2 from ID~Quantique~\cite{idqclavis2specs}. SARG04 is a four-state protocol that is equivalent to the Bennett-Brassard QKD protocol (BB84)~\cite{bennett1984} in the quantum stage. Their difference comes in the classical processing stage: in SARG04, the bases selections of Bob are used for coding the secret bits, unlike in BB84 where they are publicly revealed. Therefore, if Eve surreptitiously gets information about Bob's bases selections at any time, she can compromise the security of the QKD system running SARG04. (In contrast, a Trojan-horse attack on Bob running the BB84 protocol is normally useless \cite{vakhitov2001}, unless it is combined with other attacks \cite{makarov2006,qi2007,lydersen2010}.)

\begin{figure}
\centering
\includegraphics[width=0.66\linewidth]{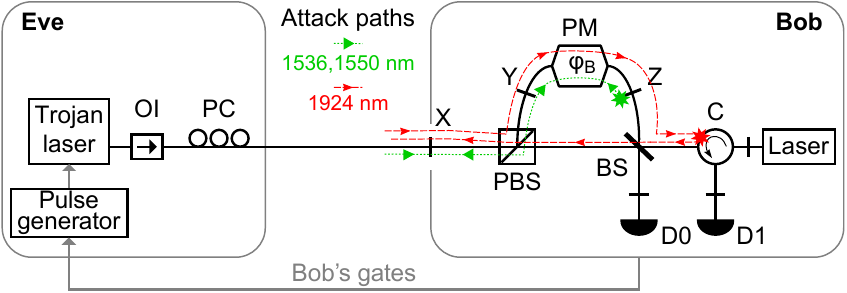}
\caption{Basic experimental schematic and attack paths at $\lambda_s = 1536~\nano\meter$ and $\lambda_l = 1924~\nano\meter$. The scheme and operation of Bob's setup is described in detail in refs.~\citen{idqclavis2specs,stucki2002}. The stars indicate the back-reflection sources exploited in ref.~\citen{jain2014} and in this work. Trojan laser models: Eblana Photonics EP1925-DM-B06-FA at $\lambda_l$ and Alcatel 1905 LMI at $\lambda_s$. OI, optical isolator; PC, polarization controller; PBS, polarizing beamsplitter; BS, 50$\,:\,$50 beamsplitter; C, circulator; D, single-photon detectors; X, Y, Z, bulkhead fiber-optic connectors.  	
\label{fig:setup}}
\end{figure}

In the attack demonstration \cite{jain2014}, it was shown that getting the bases' information in a remote manner was indeed possible via homodyne measurement of the back-reflected photons. The path taken by these photons at $1550~\nano\meter$, as depicted by the green dotted line in \cref{fig:setup}, traverses Bob's phase modulator (PM) twice. The homodyne measurement thus allowed discerning the phase applied by Bob, which is equivalent to knowing his basis selection. This `phase readout' was accurate in $>90\%$ cases even when the mean photon number of the back-reflected pulses was $\approx 3$.

Despite that, an overall attack on the QKD system did not have a chance to succeed owing to a side effect produced when the bright pulses went on to hit the detectors D0 and D1, as may be visualized in ~\cref{fig:setup}. To elaborate, the bright pulses result in a severe afterpulsing in these InGaAs/InP single-photon detectors (SPDs), which are operated in a gated mode. For a single bright pulse that hits D1, even if well outside a gate, the cumulative probability of a spurious detection event due to afterpulsing crosses $40\%$ (which is $\sim 4$ times the detection probability of a single photon) in just 5 gate periods~\cite{wiechers2011}. The resulting detection events (clicks) are accidental, i.e., erroneous in half of the cases. Hence, only a handful of Trojan-horse pulses (THPs) suffice to rapidly elevate the number of erroneous clicks and make the QBER surpass $Q_{\rm abort}$, even though Eve's actual knowledge $I^{\rm act}_E$ of the key is still quite small. An elaborate attack strategy to improve $I^{\rm act}_E$ was proposed and numerically simulated, however, it could also not simultaneously satisfy $Q \leq Q_{\rm abort}$ together with $I^{\rm act}_E > I^{\rm est}_E$, where $I^{\rm est}_E$ is the estimated (theoretical) security bound on Eve's knowledge that Clavis2 uses to produce the final secret key~\cite{jain2014}. While ref.~\citen{jain2014} did not prove that a better attack could not be constructed, the attack proposed failed in practice by a large margin.

In this Article, we provide experimental evidence that this Trojan-horse attack could however succeed if Eve were to craft bright pulses at a wavelength where the afterpulsing experienced by the SPDs is considerably lower. The underlying physics is that photons with energy lower than the bandgap of the SPD absorption layer material (InGaAs) mostly pass the material unabsorbed, thereby causing negligible afterpulsing. Indeed, we confirm experimentally that at a relatively longer wavelength $\lambda_l = 1924~\nano\meter$ the SPD has much less afterpulsing than at $\lambda_s = 1536~\nano\meter$ (similar to the wavelength used in ref.~\citen{jain2014}). We then perform a numerical comparison of the attack conditions and performance at $\lambda_l$ with these at $\lambda_s$. By means of an optimized simulation that assumes fairly realistic conditions, we show that the actual attack at $\lambda_l$ can break the security of Clavis2. The attack in itself is general enough to be potentially applicable to most discrete-variable QKD systems, and can be categorized with those that exploit vulnerabilities arising from the wavelength-dependence of optical components~\cite{li2011a,jain2015}.

\section*{Experiment}\label{sec:ExptDescpn}
While using $\lambda_l = 1924~\nano\meter$ for the attack offers the benefit of reduced afterpulsing, the transmittance and reflectance properties of different optical components inside Bob vary greatly in comparison with those measured at $\lambda_s = 1536~\nano\meter$. Most relevant to the attack, the attenuation is generally higher; for instance, the optical loss through the PM at $\lambda_l$ is $\gtrsim 20~\deci\bel$ higher than that at $\lambda_s$. 
Furthermore, the modulation itself varies with $\lambda$ since the modulator's half-wave voltage is a function of wavelength. If Eve uses light at $\lambda_l$ to estimate Bob's randomly modulated phase ($\varphi_B = 0$ or $\pi/2$ at $\lambda_s$) through the homodyne measurement of a pulse that made a single pass through the PM, the measurement outcomes will not be on orthogonal quadratures. 

Altogether, it is thus likely that compared to ref.~\citen{jain2014}, Eve would not only need to inject a larger mean photon number $\mu_{E\rightarrow B}$ into Bob, but may also require a higher mean photon number $\mu_{B\rightarrow E}$ in the back-reflection for successful homodyne measurements. To calculate the efficacy of the attack, we experimentally quantify at $\lambda_l$ (relative to $\lambda_s$) the following three aspects: increased attenuation, altered phase modulation, and decreased afterpulsing. \Cref{fig:setup} shows a schematic of the experimental setup used for various measurements.

\subsection*{Increased attenuation}\label{sec:incatt}
To gauge the increase in attenuation, we measured the optical loss of various components of Bob at both $\lambda_s$ and $\lambda_l$. In \cref{fig:setup}, the dotted line (path X--Y--$\rm Z^{\star}$--Y--X, where $^\star$ indicates the source of reflection) shows the attack path used in ref.~\citen{jain2014}. Relevant loss values are given in the left column of \cref{tab:loss_table}. With a round trip loss of $L_{\text{X--Y--}Z^{\star}\text{--Y--X}}(\lambda_s) = 2~L_{\text{X--Y}}(\lambda_s) + \Gamma_{\rm Z^{\star}} + 2~L_{\text{Y--Z}}(\lambda_s) = 58.7~\deci\bel$, Trojan-horse pulses injected with $\mu_{E\rightarrow B} \approx 2 \times 10^{6}$ photons yielded $\mu_{B\rightarrow E} \approx 4$ photons in the back-reflection from Bob. Here, $\Gamma_{\rm Z^{\star}} = 51.7~\deci\bel$ is the loss during reflection at Z, the fiber connector after Bob's PM.

\begin{table}
\centering
\caption{Comparison of optical losses in Bob at $\lambda_s$ versus $\lambda_l$. See \cref{fig:setup} for location of the paths and points. The loss during reflection $\Gamma_{\rm Z^{\star}}$ was measured at $1550~\nano\meter$ \cite{jain2014}, which we consider to be close enough to our $\lambda_s = 1536~\nano\meter$.
\label{tab:loss_table}}
{\normalsize
\begin{tabular}{l l p{4.7cm} } 
\hline
\textbf{Paths \& points} & \textbf{Loss at $\boldsymbol{\lambda_s}$} \textbf{(dB)} & \textbf{Loss at $\boldsymbol{\lambda_l}$ (dB)} \\
\hline 
X--Y & $0.9$ & $3.6$ \\ 
Y--Z & $2.6$ & $23.0$ \\
$\rm Z^{\star}$ & $51.7$  & \\ 
Z--$\rm C^{\star}$--X & & $58.4$ to $65.8$ \newline (polarization-dependent) \\
X--D0 & $8.8$ (via long arm) & $15.5$ (via short arm) \\
X--C--D1 & $9.2$ (via long arm) & $25.8$ (via short arm) \\ 
\hline    
\end{tabular}
}
\end{table}

For an attack at $\lambda_l$ with Trojan-horse pulses traversing the same path, the round trip loss would be $L_{\text{X--Y--}Z^{\star}\text{--Y--X}}(\lambda_l) = 104.9~\deci\bel$ (with the further assumption that $\Gamma_{\rm Z^{\star}}$ is independent of wavelength). The attack pulses at $\lambda_l$ would therefore face $46.2~\deci\bel$ more attenuation than at $\lambda_s$. A major contribution to this large attenuation is from the PM, which even gets doubled since the THPs travel through the PM twice.

However, since a single pass can also yield information about $\varphi_B$, Eve can opt for a different route where only either the input forward-traveling THP or the back-reflected pulse passes through Bob's PM. All Eve requires is a reasonably large source of reflection from any component after the 50$\,:\,$50 beamsplitter (BS). Indeed, during our loss measurements at $\lambda_l$ we observed a large attenuation through the optical circulator (C), a part of which stems from a rather generous back-reflection. We estimated the loss $L_{\text{Z--}\rm C^{\star}\text{--X}}(\lambda_l)$ for the path Z--$\rm C^{\star}$--X (via BS twice and polarizing beamsplitter once) using a photon-counting method, described below. 

We temporarily connected the polarization-controlled output of the $1924~\nano\meter$ laser at Z to send light towards the BS. The average power of the pulsed laser, operated at $5~\mega\hertz$ repetition rate, was $P_{\rm avg} = 21.55~\micro\watt$, corresponding to a mean photon number per pulse $\mu_{Z} = 4.14\times 10^{7}$. An SPD was connected at X to detect the back-reflections from C. To prevent other back-reflections from contributing to the photon counts, Bob's laser and detectors D0 and D1 were disconnected, and the patchcords (with open connectors) were coiled on a pencil to strongly attenuate the propagating light.

Two counters (Stanford Research Systems SR620) were used to measure   the number of optical pulses sent by the laser $N = 4.98 \times 10^6$ and the number of pulses received by the detector $n=323$ maximized over input polarization at Z. The mean photon number per pulse at X was estimated as $\mu_{X} \approx 59.7$ from the relation,  
\begin{equation}
  \frac{n-d}{N} = 1 - e^{-\mu_X \eta_d} \approx \mu_X~\eta_D, 
\label{eq:poisson}
\end{equation}
where $d=60$ is the number of dark counts and $\eta_D=8.85 \times 10^{-7}$ is the single-photon detection efficiency at $\lambda_l$, which was estimated in a separate experiment similar to the one in ref.~\citen{jain2015}. The ratio of the mean photon numbers $\mu_{Z}/\mu_{X}$ provides the overall loss $L_{\text{Z--}\rm C^{\star}\text{--X}}(\lambda_l) \approx 58.4~\deci\bel$. The dashed line in \cref{fig:setup} shows the complete attack path. Eve's THPs from the quantum channel enter the long arm of Bob, pass through the modulator, and after a reflection from the BS, propagate to the circulator. Here, they get back-reflected and then take the short arm to exit Bob, passing through the BS again. Using \cref{tab:loss_table}, this path can be characterized by a total loss $L_{\text{X--Y--Z--}\rm C^{\star}\text{--X}}(\lambda_l) = L_{\text{X--Y}}(\lambda_l) + L_{\text{Y--Z}}(\lambda_l) + L_{\text{Z--}\rm C^{\star}\text{--X}}(\lambda_l) = 85.0~\deci\bel$.

As noted above, the value of $\mu_{X}$ was polarization-sensitive. For the worst input polarization, $\mu_{X}$ decreased by $7.4~\deci\bel$, changing the overall loss to $L_{\text{X--Y--Z--}\rm C^{\star}\text{--X}}(\lambda_l) = 92.4~\deci\bel$. For the rest of the paper, we shall assume the attack pulses to be in a polarization midway between the best and the worst, leading to a loss figure of $L_{\text{X--Y--Z--}\rm C^{\star}\text{--X}}(\lambda_l) = 87.3~\deci\bel$ used to decide Eve's photon budget. In terms of photon numbers, this implies that in order to get the same number of photons out from Bob (i.e.,\ $\mu_{B\rightarrow E} \approx 4$), Eve needs to inject $\rho = 10^{(-58.7+87.3)/10} = 7.24 \times 10^{2}$ times more photons at $\lambda_l$ than at $\lambda_s$. 

\subsection*{Altered phase modulator response} \label{sec:altpmresp}
\begin{figure}
\centering
\includegraphics[width=0.5\linewidth]{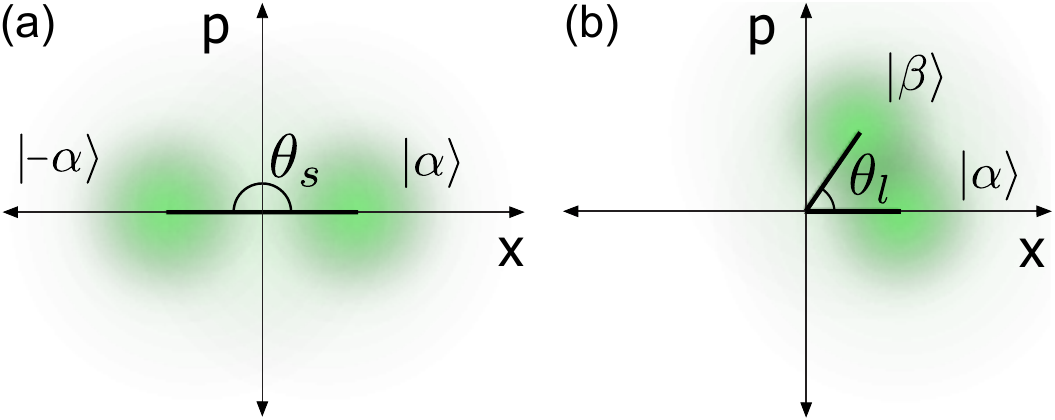}
\caption{Illustrative phase space representation of the back-reflected states. Eve attempts to discern $\varphi_B = 0$ or $\pi/2$ by performing optimal detection on the back-reflected weak coherent states $\ket{\alpha}$ and $\ket{\beta}$ that have a non-zero overlap. (a) The complex amplitude $\beta = \alpha e^{i\theta_s} = -\alpha$, as a result of the double pass at the attack wavelength of $\lambda_s$. (b) $\beta=\alpha e^{i\theta_l}$, as a result of the single pass at $\lambda_l$ through Bob's modulator. 
\label{fig:altpmresp}}
\end{figure}

We now explain an impact of the altered phase modulation experienced by Eve's THPs at $\lambda_l$ as they travel through Bob's PM. As mentioned before, Bob randomly chooses between voltages $V_0 (= 0~\volt)$ or $V_{\pi/2}$ to apply a phase $\varphi_B = 0$ or $\pi/2$ on Alice's incoming quantum signal at (or in the vicinity of) $\lambda_s = 1536~\nano\meter$. Eve's objective is to learn $\varphi_B$. The double pass through the PM in ref.~\citen{jain2014} implied that Eve had to discriminate between a pair of coherent states with angle $\theta(\lambda_s) \equiv \theta_s = 2 \times \pi/2 = \pi$ between them, as illustrated in \cref{fig:altpmresp}(a). 
At $\lambda_l = 1924~\nano\meter$, the phase modulator is expected to lose efficiency and provide less phase shift at the same voltage. Furthermore, Eve's THP only traverses it once. Assuming a linear response of the PM, one can calculate the angle $\theta_l = [V_{\pi/2}(\lambda_s)/V_{\pi/2}(\lambda_l)] \times \pi/2$ between the coherent states available to Eve. 

Since the half-wave voltage of the PM at $1924~\nano\meter$ was not specified by the manufacturer, we experimentally measured it. We constructed a balanced fiber-optic Mach-Zehnder interferometer, incorporating the path X--Z (\cref{fig:setup}) into one of its arms. We applied a square modulation voltage to the PM, and observed interference fringes at the output port of the interferometer. We adjusted the voltage amplitude until it was causing no light modulation at the output port, indicating an exact $2\pi$ phase shift. From this, we found that $V_{\pi/2}(\lambda_l) = 5.7~\volt$. By the same method with the $1536~\nano\meter$ laser, we found $V_{\pi/2}(\lambda_s) = 3.35~\volt$.

From this measurement, we calculated $\theta_l \approx 0.294\pi < \theta_s$. The increased overlap between the two states $\ket{\alpha}$ and $\ket{\beta}$ with $|\alpha|=|\beta|$, as depicted in \cref{fig:altpmresp}(b), would make discrimination between Bob's choices of $\varphi_B$ more difficult. Eve can however increase the brightness of the injected Trojan-horse pulse: this would elicit a higher mean photon number in the back-reflection, effectively translating the states farther from the origin to diminish the overlap. The increment factor that makes the distance between the states at $\lambda_l$ equal to that at $\lambda_s$ is given by
\begin{equation}
  \nu = \frac{|\alpha - \beta|^2 \text{ at } \lambda_s}{|\alpha - \beta|^2 \text{ at } \lambda_l} = \frac{1-\cos \theta_s}{1-\cos \theta_l} = 5.04, 
\label{eq:modResp}
\end{equation}
implying that a mean photon number $\mu_{B\rightarrow E} \approx 20$ at $\lambda_l$ would ensure a close-to-unity probability in the phase readout~\cite{jain2014}. 

\subsection*{Decreased afterpulse probability} \label{sec:decafpp}
To quantify the decrease in the afterpulse probabilities in Bob's detectors, we used the setup shown in \cref{fig:setup}. A single THP was synchronized to the first in a series of detection gates~\cite{wiechers2011,jain2014} of Bob, and the times at which clicks occurred in the onward gates were then recorded. The delay of the THP relative to the first gate was adjusted such that the pulses going through Bob's long arm hit the detectors just a few nanoseconds after the gate was applied by Bob. Although we did utilize a polarization controller, only a maximum of $\sim 45\%$ of the incoming optical power at $\lambda_l$ could be routed through the long arm. The remaining light, after having suffered propagation losses through the short arm, hit D0 and D1 around $50~\nano\second$ \emph{before} the first gate (propagation time through the short arm is $\approx 50~\nano\second$ faster than the long arm in Clavis2~\cite{stucki2002}). These light pulses before the gate were found to be the dominant cause for increased noise in the detectors.

\begin{figure}
\centering
\includegraphics[width=0.6\linewidth]{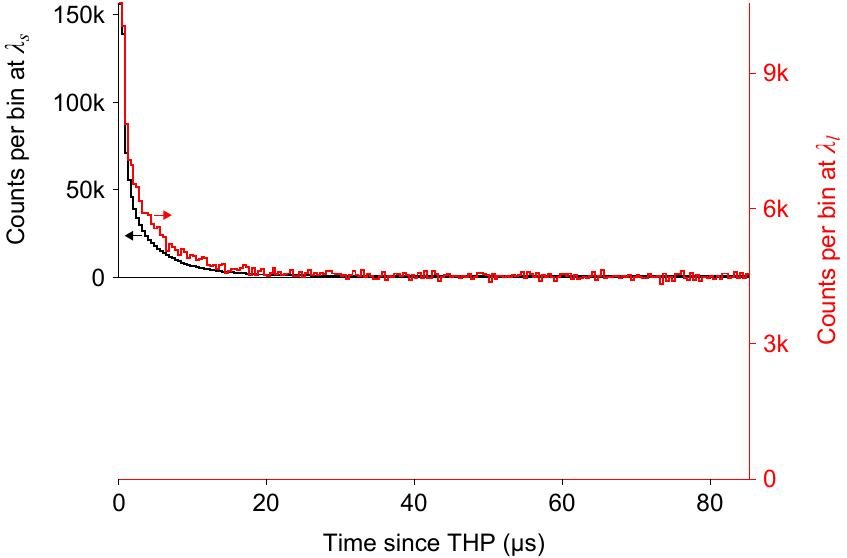}
\caption{Afterpulse profiles at $\lambda_s = 1536~\nano\meter$ and $\lambda_l = 1924~\nano\meter$. Note that the histograms are rescaled such that their peak counts and dark count rates match in the plot, making visual comparison of decay curves easy. The decay curves are similar but not identical. A total of $10^6$ counts were histogrammed at each wavelength. The originally collected histogram data exhibited a saturation effect, in which count rate in later bins was slightly suppressed (by $6.4\%$ for $\lambda_s$, $1.0\%$ for $\lambda_l$) because of significant click probability in early bins. This has been corrected in the plotted histograms, increasing their total count number above $10^6$.
\label{fig:profileAfp}}
\end{figure}

\Cref{fig:profileAfp} shows the time distribution of counts recorded in detector D0 at the wavelengths $\lambda_s$ and $\lambda_l$. Each of the histograms was prepared by recording $10^6$ counts. To make the most of the limited number of histogram bins in the counter (SR620), each bin was $0.4~\micro\second$ wide and included counts from two consecutive gates. This allowed us to cover a time range of $> 80~\micro\second$. THPs with mean photon numbers $\mu_s = 2.68 \times 10^4$  and $\mu_l = 8.32 \times 10^{7}$ were used for wavelengths $\lambda_s$ and $\lambda_l$ respectively. Despite $\mu_s \ll \mu_l$, the data acquisition for the latter took much longer, indicating that most of the clicks were actually (thermal) dark counts. The number of counts per bin settled down at a constant value, representing dark counts, after $\sim 40~\micro\second$ (right half of the histogram). The total number of thermal dark counts collected could then be calculated by multiplying this value by the total number of bins in the entire histogram. All remaining counts could then be attributed to afterpulsing. \Cref{tab:counts} lists these counts at the two wavelengths. The afterpulse counts ($ApC$) make the bulk of the counts at $\lambda_s$, while dark counts ($DC$) are in the majority at $\lambda_l$.

It can also be observed in \cref{fig:profileAfp} that afterpulsing decay profile at both wavelengths is roughly similar, however the ratio of longer to shorter lifetime components is slightly larger at $\lambda_l$. Although this would help our modeled attack \cite{jain2014}, for simplicity we have conservatively assumed that the decay parameters at $\lambda_l$ are the same as at $\lambda_s$ \cite{wiechers2011,endnote20160902}, aside from different overall afterpulse probability.

\begin{table}
	\caption{Counts due to thermal dark noise ($DC$) and afterpulsing ($ApC$), extracted from \cref{fig:profileAfp} and corrected for the saturation effect. $(ApC + DC)$ is greater than $10^6$ owing to this correction.
		\label{tab:counts}}
	\centering
	\begin{tabular}{ p{1.6cm} p{2.2cm} p{1.6cm}  p{1.4cm} }
		\hline
		\textbf{$\boldsymbol{\lambda}$ (nm)} & $\boldsymbol{\mu}$ & $\boldsymbol{ApC}$ & $\boldsymbol{DC}$ \\ \hline
		$1536$   & $2.68 \times 10^4$ & $867760$ & $162854$ \\
		$1924$   & $8.32 \times 10^7$ & ~~$44981$ & $962140$ \\ \hline
	\end{tabular}
\end{table}

To compute a numerical factor $\gamma$ that compares the afterpulsing noise induced at the two wavelengths, we first take the ratio $(ApC/DC)$ at each wavelength. Then, assuming the dark count probability per detector gate stayed constant between the two measurements, we take a ratio of these ratios. We assume a linear scaling of the afterpulse probability with the energy of the THP, and further normalise for the dissimilar mean photon numbers $\mu_s$ and $\mu_l$ of the THPs. The numerical factor is then
\begin{equation}
  \gamma = \frac{\mu_s}{\mu_l} \frac{({ApC_l}/{DC_l})}{({ApC_s}/{DC_s})} = 2.83 \times 10^{-6}.
\label{eq:redafp}
\end{equation}
In other words, a photon at $\lambda_l$ is only $2.83 \times 10^{-6}$ times as likely to cause an afterpulse as a photon at $\lambda_s$. 

\section*{Attack modeling and discussion}
Relative to $\lambda_s$, an attack at $\lambda_l$ can thus effectively decrease the afterpulsing probability in D0 by
\begin{equation}
 \delta_0 = \rho \nu \gamma = 1.03 \times 10^{-2}.
\label{eq:delta0}
\end{equation}
The factor $\rho \nu = 3.65 \times 10^{3}$ combines the results discussed previously on the aspects of increased attenuation and altered phase modulation, which required THPs injected into Bob at $\lambda_l$ to be $\rho \nu$ times brighter than at $\lambda_s$ to ensure optimal attack performance. 

To calculate the afterpulsing probability for D1, one must also consider different losses from Bob's entrance to detectors D0 and D1 for the two attack paths (via the long arm at $\lambda_s$ and short arm at $\lambda_l$, as shown in \cref{fig:setup}). We minimised $L_{\text{X--Y}}(\lambda_l)$ by adjusting input polarisation at X, then measured losses between X and the detectors through the short arm. $L_{\text{X--C--D1}}(\lambda_l)$ varied by a factor of 11 over the input polarization, while $L_{\text{X--D0}}(\lambda_l)$ unexpectedly was independent of the input polarization. Using the measured loss values (listed in the last two rows in \cref{tab:loss_table}), we calculate the effective decrease in the afterpulsing probability in D1
\begin{align}
 \delta_1 &= \delta_0 \times 10^{[L_{\text{X--C--D1}}(\lambda_s)-L_{\text{X--D0}}(\lambda_s)-L_{\text{X--C--D1}}(\lambda_l)+L_{\text{X--D0}}(\lambda_l)]/10}\nonumber \\ &= 1.05 \times 10^{-3}.
\label{eq:delta1}
\end{align}

With afterpulsing amplitudes reduced by $\delta_0$ and $\delta_1$, we have repeated the simulation of the attack strategy proposed in ref.~\citen{jain2014}. Let us first recap this strategy, in which Eve manipulates packets or `frames' \cite{idqclavis2specs} of quantum signals traveling from Alice to Bob in the quantum channel. For instance, she may simply block the quantum signals for several contiguous time slots in a frame, thereby preventing any detection clicks (except those arising from dark counts) in Bob over a certain period of time. Conversely, she could substitute the quantum channel with a low-loss version to increase the detection probability in another group of slots. Such actions provide Eve some control over when Bob's SPDs enter deadtime~\cite{endnote20160927} inside the frame, which is used to increase the efficacy of her attack. This is essentially done by attacking in bursts, i.e.,\ probing the phase modulator by sending bright THPs in a group of slots, thus making the SPDs enter deadtime as quickly as possible to let the afterpulses decay harmlessly and contribute as little as possible to the QBER. By balancing the usage of the low-loss line and the number of slots blocked per frame, Eve can also ensure that Bob does not notice any significant deviation of the observed detection rate (typically averaged over a large number of frames). 

A numerical simulation modeling the above attack strategy during the operation of the QKD protocol is used to calculate Bob's incurred QBER $Q$ and Eve's actual knowledge of the raw key $I^{\rm act}_E$. This is performed for different attack combinations, i.e.,\ by varying the number of slots that are blocked or simply passed via the low-loss line (with or without accompanying THPs). If for at least one combination, $I^{\rm act}_E$ exceeds the estimation $I^{\rm est}_E$ from the security proof but $Q < Q_{\rm abort}$, the attack strategy is successful in breaching the security. 

For an attack at $\lambda_l$, we have been able to find several such combinations for the given frame size of $N_f = 1075$ slots and a quantum channel transmittance $T=0.25$. For instance, in one such combination, a total of $433$ slots out of $N_f$ are blocked by Eve. The remaining $642$ slots pass from Alice to Bob via a low-loss line with transmittance $T_{LL} = 0.5$, and out of them only $334$ slots --- periodically distributed in $12$ bursts of $28$ slots each inside the frame --- are accompanied by THPs to read the modulation. With this attack combination, we were able to obtain $I^{\rm act}_E = 0.515 > I^{\rm est}_E = 0.506$ (calculation based on Clavis2 parameters and the attack conditions~\cite{jain2014}) and $Q = 7.8\% < Q_{\rm abort} \approx 8\%$ (empirically determined in ref.~\citen{jain2011}). We remark here that for a similar value of $Q$, the best optimized attacks at $\lambda_s$ could not even yield $I^{\rm act}_E \sim 0.080$. Furthermore, in contrast to the $T_{LL}=0.9$ used in ref.~\citen{jain2014}, implementing the attack strategy with $T_{LL} = 0.5$ here makes the attack closer to be feasible in practice. 

Note that in the simulation, we have mixed measurement results from two samples of Clavis2 system. The optical loss measurements at $\lambda_l$ and the relative decrease in afterpulsing come from the system installed in Waterloo (Bob module serial number 08020F130), while the decay parameters of trap levels in avalanche photodiodes measured at $\lambda_s$ come from the system in Erlangen (Bob module serial number 08008F130) \cite{endnote20160902}. We further note that the latter figures vary significantly between D0 and D1, although the two avalanche photodiodes were of the same type and at the same temperature \cite{wiechers2011}. Therefore our simulation only gives a rough indication of attack performance. Results of the actual attack, if it is performed, will vary from sample to sample. However, also note that we have tested a single long wavelength of $1924~\nano\meter$; a different wavelength may well yield better attack performance. Finally, more recent commercial systems deploy SPDs with much better efficiencies and afterpulsing characteristics and, as noted in ref.~\citen{jain2014}, this benefits the eavesdropping strategy.

We expect homodyne detection at $1924~\nano\meter$ to be easy to implement by using p-i-n diodes with extended infrared response~\cite{InGaAsPINdiodesIG22Series,InGaAsPINdiodesG12182series}. Based on the published specs, the latter should provide detection performance in our setting similar to that demonstrated at $1550~\nano\meter$ \cite{jain2014}. Separating Eve from Bob by some distance of fiber does not degrade the attack very fast; we have measured $7.5~\deci\bel\per\kilo\meter$ loss at $1924~\nano\meter$ in a $16.5~\centi\meter$ diameter spool of Corning SMF-28e \cite{smf28e} fiber.

The easiest countermeasure to protect the QKD system from this attack is to properly filter the light entering the system~\cite{jain2015,Lucamarini2015}. E.g.,\ adding a narrow-pass filter at Bob's entrance will force Eve to use the signal wavelength $\lambda_s$ and reduce her attack performance to the original failure, provided poor detector afterpulsing properties are maintained in production~\cite{jain2014}. Another countermeasure would be to use a QKD protocol that does not require the receiver's PM settings to be secret, such as BB84 with decoy states~\cite{vakhitov2001,hwang2003,scarani2009}. However, protecting the source's PM settings will still be required in most QKD protocols \cite{Lucamarini2015,sajeed2015}.

\section*{Conclusion}

In conclusion, we have shown that despite the increased attenuation and sub-optimal phase modulation experienced around $1924~\nano\meter$, the Trojan-horse attack performed at this wavelength has a very good chance of being invisible, because the afterpulsing experienced by Bob's detectors is extremely low. This attack is mostly implementable with commercial off-the-shelf components. Therefore, an urgent need exists to incorporate effective countermeasures into practical QKD systems to thwart such threats.


\begin{thebibliography}{10}
	\expandafter\ifx\csname url\endcsname\relax
	\def\url#1{\texttt{#1}}\fi
	\expandafter\ifx\csname urlprefix\endcsname\relax\def\urlprefix{URL }\fi
	\providecommand{\bibinfo}[2]{#2}
	\providecommand{\eprint}[2][]{\url{#2}}
	
	\bibitem{bennett1984}
	\bibinfo{author}{Bennett, C.~H.} \& \bibinfo{author}{Brassard, G.}
	\newblock \bibinfo{title}{Quantum cryptography: Public key distribution and
		coin tossing}.
	\newblock In \emph{\bibinfo{booktitle}{Proc. IEEE International Conference on
			Computers, Systems, and Signal Processing (Bangalore, India)}},
	\bibinfo{pages}{175--179} (\bibinfo{publisher}{IEEE Press},
	\bibinfo{address}{New York}, \bibinfo{year}{1984}).
	
	\bibitem{gisin2002}
	\bibinfo{author}{Gisin, N.}, \bibinfo{author}{Ribordy, G.},
	\bibinfo{author}{Tittel, W.} \& \bibinfo{author}{Zbinden, H.}
	\newblock \bibinfo{title}{Quantum cryptography}.
	\newblock \emph{\bibinfo{journal}{Rev. Mod. Phys.}}
	\textbf{\bibinfo{volume}{74}}, \bibinfo{pages}{145--195}
	(\bibinfo{year}{2002}).
	
	\bibitem{scarani2009}
	\bibinfo{author}{Scarani, V.} \emph{et~al.}
	\newblock \bibinfo{title}{The security of practical quantum key distribution}.
	\newblock \emph{\bibinfo{journal}{Rev. Mod. Phys.}}
	\textbf{\bibinfo{volume}{81}}, \bibinfo{pages}{1301--1350}
	(\bibinfo{year}{2009}).
	
	\bibitem{makarov2011}
	\bibinfo{author}{Makarov, V.}
	\newblock \bibinfo{title}{Cracking quantum cryptography}.
	\newblock In \emph{\bibinfo{booktitle}{CLEO/Europe and EQEC 2011 Conference
			Digest}}, \bibinfo{pages}{$\textrm{ED3\_1}$} (\bibinfo{publisher}{Optical
		Society of America}, \bibinfo{year}{2011}).
	
	\bibitem{Scarani2014}
	\bibinfo{author}{Scarani, V.} \& \bibinfo{author}{Kurtsiefer, C.}
	\newblock \bibinfo{title}{The black paper of quantum cryptography: real
		implementation problems}.
	\newblock \emph{\bibinfo{journal}{Theor. Comput. Sci.}}
	\textbf{\bibinfo{volume}{560}}, \bibinfo{pages}{27--32}
	(\bibinfo{year}{2014}).
	
	\bibitem{Lo2014}
	\bibinfo{author}{Lo, H.-K.}, \bibinfo{author}{Curty, M.} \&
	\bibinfo{author}{Tamaki, K.}
	\newblock \bibinfo{title}{{Secure quantum key distribution}}.
	\newblock \emph{\bibinfo{journal}{Nat. Photonics}}
	\textbf{\bibinfo{volume}{8}}, \bibinfo{pages}{595--604}
	(\bibinfo{year}{2014}).
	
	\bibitem{jain2016}
	\bibinfo{author}{Jain, N.} \emph{et~al.}
	\newblock \bibinfo{title}{Attacks on practical quantum key distribution systems
		(and how to prevent them)}.
	\newblock \emph{\bibinfo{journal}{Contemp. Phys.}}
	\textbf{\bibinfo{volume}{57}}, \bibinfo{pages}{366--387}
	(\bibinfo{year}{2016}).
	
	\bibitem{liang2014}
	\bibinfo{author}{Liang, L.-M.}, \bibinfo{author}{Sun, S.-H.},
	\bibinfo{author}{Jiang, M.-S.} \& \bibinfo{author}{Li, C.-Y.}
	\newblock \bibinfo{title}{Security analysis on some experimental quantum key
		distribution systems with imperfect optical and electrical devices}.
	\newblock \emph{\bibinfo{journal}{Front. Phys.}} \textbf{\bibinfo{volume}{9}},
	\bibinfo{pages}{613--628} (\bibinfo{year}{2014}).
	
	\bibitem{gisin2006}
	\bibinfo{author}{Gisin, N.}, \bibinfo{author}{Fasel, S.},
	\bibinfo{author}{Kraus, B.}, \bibinfo{author}{Zbinden, H.} \&
	\bibinfo{author}{Ribordy, G.}
	\newblock \bibinfo{title}{Trojan-horse attacks on quantum-key-distribution
		systems}.
	\newblock \emph{\bibinfo{journal}{Phys. Rev. A}} \textbf{\bibinfo{volume}{73}},
	\bibinfo{pages}{022320} (\bibinfo{year}{2006}).
	
	\bibitem{vakhitov2001}
	\bibinfo{author}{Vakhitov, A.}, \bibinfo{author}{Makarov, V.} \&
	\bibinfo{author}{Hjelme, D.~R.}
	\newblock \bibinfo{title}{Large pulse attack as a method of conventional
		optical eavesdropping in quantum cryptography}.
	\newblock \emph{\bibinfo{journal}{J. Mod. Opt.}} \textbf{\bibinfo{volume}{48}},
	\bibinfo{pages}{2023--2038} (\bibinfo{year}{2001}).
	
	\bibitem{jain2014}
	\bibinfo{author}{Jain, N.} \emph{et~al.}
	\newblock \bibinfo{title}{Trojan-horse attacks threaten the security of
		practical quantum cryptography}.
	\newblock \emph{\bibinfo{journal}{New J. Phys.}} \textbf{\bibinfo{volume}{16}},
	\bibinfo{pages}{123030} (\bibinfo{year}{2014}).
	
	\bibitem{scarani2004}
	\bibinfo{author}{Scarani, V.}, \bibinfo{author}{Ac\'{i}n, A.},
	\bibinfo{author}{Ribordy, G.} \& \bibinfo{author}{Gisin, N.}
	\newblock \bibinfo{title}{Quantum cryptography protocols robust against photon
		number splitting attacks for weak laser pulse implementations}.
	\newblock \emph{\bibinfo{journal}{Phys. Rev. Lett.}}
	\textbf{\bibinfo{volume}{92}}, \bibinfo{pages}{057901}
	(\bibinfo{year}{2004}).
	
	\bibitem{idqclavis2specs}
	\bibinfo{note}{{C}lavis2 specification sheet,
		\url{http://www.idquantique.com/images/stories/PDF/clavis2-quantum-key-distribution/clavis2-specs.pdf},
		visited 16 Apr 2017}.
	
	\bibitem{makarov2006}
	\bibinfo{author}{Makarov, V.}, \bibinfo{author}{Anisimov, A.} \&
	\bibinfo{author}{Skaar, J.}
	\newblock \bibinfo{title}{Effects of detector efficiency mismatch on security
		of quantum cryptosystems}.
	\newblock \emph{\bibinfo{journal}{Phys. Rev. A}} \textbf{\bibinfo{volume}{74}},
	\bibinfo{pages}{022313} (\bibinfo{year}{2006}).
	\newblock \bibinfo{note}{Erratum ibid. \textbf{78}, 019905 (2008)}.
	
	\bibitem{qi2007}
	\bibinfo{author}{Qi, B.}, \bibinfo{author}{Fung, C.-H.~F.},
	\bibinfo{author}{Lo, H.-K.} \& \bibinfo{author}{Ma, X.}
	\newblock \bibinfo{title}{Time-shift attack in practical quantum
		cryptosystems}.
	\newblock \emph{\bibinfo{journal}{Quant. Inf. Comp.}}
	\textbf{\bibinfo{volume}{7}}, \bibinfo{pages}{73--82} (\bibinfo{year}{2007}).
	
	\bibitem{lydersen2010}
	\bibinfo{author}{Lydersen, L.} \& \bibinfo{author}{Skaar, J.}
	\newblock \bibinfo{title}{Security of quantum key distribution with bit and
		basis dependent detector flaws}.
	\newblock \emph{\bibinfo{journal}{Quant. Inf. Comp.}}
	\textbf{\bibinfo{volume}{10}}, \bibinfo{pages}{60--76}
	(\bibinfo{year}{2010}).
	
	\bibitem{stucki2002}
	\bibinfo{author}{Stucki, D.}, \bibinfo{author}{Gisin, N.},
	\bibinfo{author}{Guinnard, O.}, \bibinfo{author}{Ribordy, G.} \&
	\bibinfo{author}{Zbinden, H.}
	\newblock \bibinfo{title}{Quantum key distribution over 67 km with a plug\&play
		system}.
	\newblock \emph{\bibinfo{journal}{New J. Phys.}} \textbf{\bibinfo{volume}{4}},
	\bibinfo{pages}{41} (\bibinfo{year}{2002}).
	
	\bibitem{wiechers2011}
	\bibinfo{author}{Wiechers, C.} \emph{et~al.}
	\newblock \bibinfo{title}{After-gate attack on a quantum cryptosystem}.
	\newblock \emph{\bibinfo{journal}{New J. Phys.}} \textbf{\bibinfo{volume}{13}},
	\bibinfo{pages}{013043} (\bibinfo{year}{2011}).
	
	\bibitem{li2011a}
	\bibinfo{author}{Li, H.-W.} \emph{et~al.}
	\newblock \bibinfo{title}{Attacking a practical quantum-key-distribution system
		with wavelength-dependent beam-splitter and multiwavelength sources}.
	\newblock \emph{\bibinfo{journal}{Phys. Rev. A}} \textbf{\bibinfo{volume}{84}},
	\bibinfo{pages}{062308} (\bibinfo{year}{2011}).
	
	\bibitem{jain2015}
	\bibinfo{author}{Jain, N.} \emph{et~al.}
	\newblock \bibinfo{title}{Risk analysis of {T}rojan-horse attacks on practical
		quantum key distribution systems}.
	\newblock \emph{\bibinfo{journal}{IEEE J. Sel. Top. Quantum Electron.}}
	\textbf{\bibinfo{volume}{21}}, \bibinfo{pages}{6600710}
	(\bibinfo{year}{2015}).
	
	\bibitem{endnote20160902}
	\bibinfo{note}{The decay parameters and $\rm Z^{\star}$ were measured at
		$1550~\nano\meter$ \cite{wiechers2011,jain2014}, which we consider to be
		close enough to our $\lambda_s = 1536~\nano\meter$.}
	
	\bibitem{endnote20160927}
	\bibinfo{note}{Deadtime denotes a period in which both D0 and D1 are
		insensitive to single photons and cannot register detection clicks. In
		Clavis2, a $10\,\upmu$s long deadtime is triggered by a click in either of
		the detectors \cite{wiechers2011}.}
	
	\bibitem{jain2011}
	\bibinfo{author}{Jain, N.} \emph{et~al.}
	\newblock \bibinfo{title}{Device calibration impacts security of quantum key
		distribution}.
	\newblock \emph{\bibinfo{journal}{Phys. Rev. Lett.}}
	\textbf{\bibinfo{volume}{107}}, \bibinfo{pages}{110501}
	(\bibinfo{year}{2011}).
	
	\bibitem{InGaAsPINdiodesIG22Series}
	\bibinfo{note}{Extended InGaAs PIN photodiodes IG22-series,
		\url{http://www.lasercomponents.com/us/product/ingaas-500-2600-nm-1/},
		visited 16 Apr 2017}.
	
	\bibitem{InGaAsPINdiodesG12182series}
	\bibinfo{note}{InGaAs PIN photodiodes G12182 series,
		\url{http://www.hamamatsu.com/resources/pdf/ssd/g12182_series_kird1118e.pdf},
		visited 16 Apr 2017}.
	
	\bibitem{smf28e}
	\bibinfo{note}{{C}orning SMF-28e optical fiber,
		\url{http://www.princetel.com/datasheets/SMF28e.pdf}, visited 16 Apr 2017}.
	
	\bibitem{Lucamarini2015}
	\bibinfo{author}{Lucamarini, M.} \emph{et~al.}
	\newblock \bibinfo{title}{Practical security bounds against the {T}rojan-horse
		attack in quantum key distribution}.
	\newblock \emph{\bibinfo{journal}{Phys. Rev. X}} \textbf{\bibinfo{volume}{5}},
	\bibinfo{pages}{031030} (\bibinfo{year}{2015}).
	
	\bibitem{hwang2003}
	\bibinfo{author}{Hwang, W.-Y.}
	\newblock \bibinfo{title}{Quantum key distribution with high loss: Toward
		global secure communication}.
	\newblock \emph{\bibinfo{journal}{Phys. Rev. Lett.}}
	\textbf{\bibinfo{volume}{91}}, \bibinfo{pages}{057901}
	(\bibinfo{year}{2003}).
	
	\bibitem{sajeed2015}
	\bibinfo{author}{Sajeed, S.} \emph{et~al.}
	\newblock \bibinfo{title}{Attacks exploiting deviation of mean photon number in
		quantum key distribution and coin tossing}.
	\newblock \emph{\bibinfo{journal}{Phys. Rev. A}} \textbf{\bibinfo{volume}{91}},
	\bibinfo{pages}{032326} (\bibinfo{year}{2015}).
	
\end{thebibliography}

\section*{Acknowledgements}
We thank ID~Quantique for cooperation, technical assistance, and providing us the QKD hardware. This work was funded by Industry Canada, CFI, NSERC (programs Discovery and CryptoWorks21), Ontario MRI, and the US Office of Naval Research. N.J.\ acknowledges the warm hospitality of the Institute for Quantum Computing, where this work was carried out. 

\section*{Author contributions statement}
S.S.\ and C.M.\ performed the experiments. N.J.\ performed attack modeling and contributed to the experiments. V.M.\ supervised the study. All authors performed data analysis and contributed to writing the article.

\section*{Additional information}
\textbf{Competing financial interests:} The authors declare no competing financial interests.

\end{document}